\documentclass[osajnl,preprint,showpacs,superscriptaddress,12pt]{revtex4-1} %% use 12pt for preprint option
\usepackage{amsmath,amssymb,graphicx}

\begin{document}

\title{A one-dimensional ultracold medium of extreme optical depth}

\author{Frank Blatt}
\author{Thomas Halfmann}
\author{Thorsten Peters}
\affiliation{Institut f\"ur Angewandte Physik, Technische Universit\"at Darmstadt, Hochschulstrasse 6, 64289 Darmstadt, Germany}

\begin{abstract}
We report on the preparation of a one-dimensional ultracold medium in a hollow-core photonic crystal fiber, reaching an effective optical depth of 1000(150). We achieved this extreme optical depth by transferring atoms from a magneto-optical trap into a far-detuned optical dipole trap inside the hollow-core fiber, yielding up to 2.5(3)$\times$10$^5$ atoms inside the core with a loading efficiency of 2.5(6)~\%. The preparation of an ultracold medium of such huge optical depth paves the way towards new applications in quantum optics and nonlinear optics.
\end{abstract}

%\ocis{(020.3320) Laser cooling; (020.7010) Laser trapping; (060.5295) Photonic crystal fibers; (300.6210) Spectroscopy, atomic.}

\maketitle

Guiding ultracold atoms into or through hollow-core fibers \cite{OOL93} has been subject of experimental efforts in the recent two decades of quantum optics. In such setups, atoms and light are strongly confined transversally over longitudinally macroscopic distances, while in free space transverse and longitudinal confinement cannot be controlled independently. This represents, e.g., an alternative approach towards strong light-matter interactions, which so far are typically achieved by tight confinement of light and matter in high-finesse micro-cavities \cite{WVE06}. Moreover, long-distance guiding of matter waves for, e.g., atom interferometry \cite{CSP09} becomes possible in such setups.

As a specific example for the potential of strong interactions of confined light-matter systems \cite{CGM08,AHK11,HA12,HNR12,AHC13} we note applications of stationary light pulses (SLPs) \cite{BZL03,LLP09}. The latter are a variant of electromagnetically induced transparency (EIT) \cite{FIM05} with counterpropagating control fields. SLPs permit ``trapping" of light pulses in an EIT-driven medium, similar to a cavity, but without the need for mirrors. The phenomenon triggered a number of interesting proposals for applications, e.g.,``crystallization of light" \cite{CGM08} using repulsive optical interactions to generate a Tonks-Girardeau gas \cite{T36,G60} of photons, preparation of a Luttinger liquid with photons \cite{AHK11}, implementing Sine-Gordon and Bose-Hubbard dynamics with photons \cite{HA12}, simulation of Cooper pairing \cite{HNR12}, or mimicking of relativistic theories \cite{AHC13}. A key requirement in all these proposals are one-dimensional media of extreme on-resonant opaqueness, characterized by an optical depth $OD=-\ln(I_{out}/I_{in})$ beyond 1000. So far, such media were not available.

We note, that some recent efforts in magneto-optical traps (MOTs) achieved high ODs of 150 to 300 on optical transitions relevant to light storage protocols \cite{SBH13,CLW13} and up to 1000 on a cycling transition \cite{SBH13}. Other approaches were based on hollow-core fibers filled with an atomic gas at or above room temperature \cite{SBV08,SEA13}. Here, very extreme ODs above $10^4$ were observed. In such setups, however, inhomogeneous broadening significantly reduces the effective OD by the ratio of the homogeneous to the inhomogeneous linewidth $\Gamma_{hom}/\Gamma_{inh}$ \cite{SEA13}. Thus, in the aforementioned cases the effective OD was $\sim$1200 \cite{SBV08} and $\sim$300 \cite{SEA13}, respectively. Moreover, strong transverse confinement in the range of several microns can lead to significant transit-time broadening and dephasing collisions with the fiber wall in these setups \cite{GBR06}. Loading laser-cooled atoms into a hollow-core fiber and guiding them (well separated from the fiber walls) by an optical dipole trap overcomes these problems.

First attempts applied capillaries with diameters ranging from 2~$\mu$m to more than hundred microns to guide atoms from a thermal source by a red- or blue-detuned optical dipole trap \cite{RMV95,RDC96,INS96,RZD97,DHB99,MCA00}. However, capillaries suffer easily from transverse multi-mode operation of the guiding field, attenuation of the beam while it propagates along the capillary, as well as the formation of speckles \cite{RMV95,MCA00}, which are all detrimental to guiding of atoms.
With the availability of hollow-core photonic crystal fibers (HCPCFs) \cite{R03}, also low-loss, single-mode guiding of light in hollow fibers became possible. Since then, several groups demonstrated guiding of atoms into \cite{CWS08,BHB09,BHP11} or through HCPCFs \cite{TK07,VMW10}. In these experiments the source of atoms was either a thermal oven \cite{TK07}, laser-cooled atoms from a magneto-optical trap (MOT) \cite{VMW10,BHB09,BHP11}, or even a Bose-Einstein condensate \cite{CWS08}. Apart from the latter, technically very challenging approach, the most successful preparation of a medium at large OD so far was reported by Bajcsy \textit{et al.} \cite{BHP11}. The authors demonstrated the loading of $3\times10^4$ $^{87}$Rb atoms from a MOT into a HCPCF (length of 3~cm, core diameter $\sim7~\mu$m). This corresponded to a loading efficiency of $0.3~\%$ from the MOT into the fiber. It yielded ODs of (12,60,180) on the transitions $|F=2\rangle \rightarrow |F'=1,2,3\rangle$, respectively, of the $D_2$ line.

In this work we report on loading of laser-cooled $^{87}$Rb atoms from a MOT into a HCPCF of core diameter $\sim7~\mu$m, reaching the largest loading efficiency and optical depth achieved in such experiments so far. Our setup allows for cooling of atoms until they enter the HCPCF, a feature not possible, e.g., in the work of Ref. \cite{BHP11}. 

\begin{figure}[htbp]
	\centerline{\includegraphics[width=8.4cm]{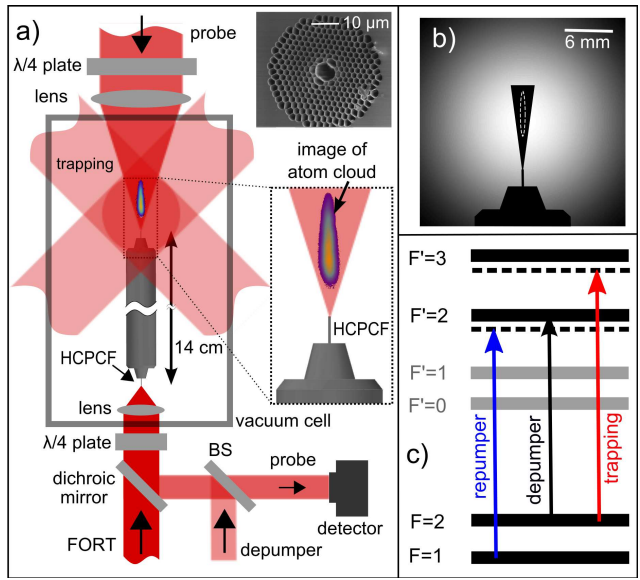}}
	\caption{(color online) (a) Schematic experimental setup and scanning electron micrograph of the HCPCF (upper right corner). BS: beam splitter. (b) Intensity profile of the dark funnel repumper beams. The dashed line indicates the size and position of the atom cloud before the transfer process into the fiber. Besides the funnel, also the region of the fiber and its mount is kept dark to avoid stray light. (c) MOT coupling scheme.}
	\label{fig:Fig1}
\end{figure}

We performed the experiment with $^{87}$Rb atoms trapped in a standard vapor cell MOT with rectangular-shaped quadrupole field coils \cite{PWB12}. 
The three individual trapping beams were detuned by -18~MHz from the cycling transition with a total power of 38~mW and full-width-at-half-maximum (FWHM) diameters of 20~mm. The repumper, detuned by -6~MHz from the transition $|F=1\rangle\rightarrow|F'=2\rangle$, had the same diameter and a power up to 3~mW. With this setup we typically trapped 10$^7$ atoms at a temperature of $T\simeq120~\mu$K in the MOT after a loading period of 1~s. The HCPCF (HC-800-02, NKT Photonics, length 14 cm, core diameter (5.5-7.0)~$\mu$m, transmission window 770~nm - 870~nm, see Fig.~\ref{fig:Fig1}(a)) was mounted vertically inside the vacuum system. The distance of the upper fiber tip from the center of the MOT was about 5.5~mm. We prepared a circularly polarized, red-detuned far off-resonant optical trap (FORT) inside the HCPCF to guide atoms into the fiber core and prevent collisions with the room temperature fiber wall. The radiation of the FORT was generated in a home-built tapered amplifier system, operating at a wavelength of 855~nm and providing an output power up to 1~W. After passing through several optical elements, including a two-stage Faraday isolator and a single-mode fiber, around 270~mW of this power could be coupled into the Gaussian mode of the HCPCF with a coupling efficiency above 90~$\%$. This corresponds to a trapping potential depth up to 5~mK and a transverse trap frequency of 80~kHz. As the FORT quickly diverges after leaving the fiber (numerical aperture $~0.1$), the potential required to trap atoms from the MOT extends only to a distance of roughly $200~\mu$m from the fiber tip.  In order to transfer the atoms from the MOT into the HCPCF most efficiently, the relatively large atom cloud ($2.5 \times 0.4^2$ mm$^3$ 1/e half widths) above the fiber must hence be cooled, compressed and shifted as close as possible to the fiber tip. 

The driving laser beams for probing and guiding atoms inside the HCPCF were focused into the fiber with an aspheric doublet lens ($f=75$~mm) located outside above the vacuum cell and with an aspheric lens ($f=7.5$~mm) located below the fiber inside the vacuum system (see Fig.~\ref{fig:Fig1}(a)). 
\begin{figure*}[htbp]
	\centerline{\includegraphics[width=17.5cm]{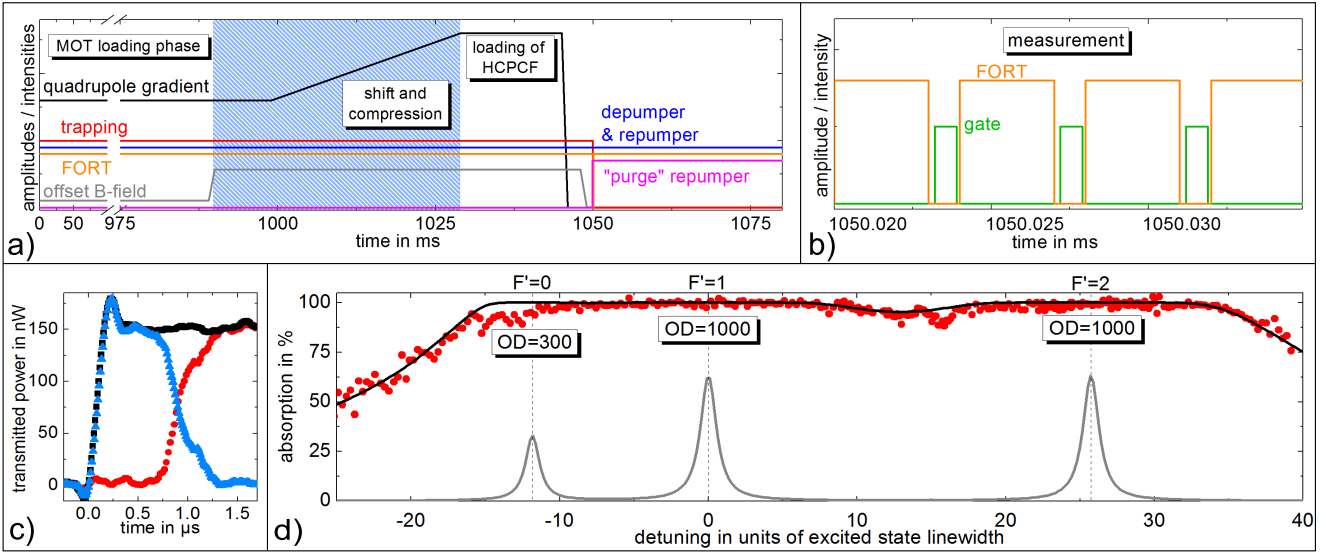}}
	\caption{(color online) (a) Timing sequence of the loading procedure. (b) Absorption measurement sequence. (c) Absolute transmitted probe power through the HCPCF vs. time without (black squares) and with (red circles) atoms loaded into the fiber. The data was averaged over 8 loading cycles. The blue triangles depict the difference of both signals. (c) Absorption spectrum on the transitions $|F=1\rangle \rightarrow |F'=0,1,2\rangle$ vs. probe laser detuning from state $|F'=1\rangle$ (symbols). The black line shows a calculated spectrum for $OD=(300(45), 1000(150), 1000(150))$ on the transitions $|F=1\rangle \rightarrow |F'=0,1,2\rangle$. The gray line shows a spectrum for $OD\simeq1$ for reference.}
	\label{fig:Fig2}
\end{figure*}
We employed the following strategy to load the HCPCF: We applied the dark SPOT technique \cite{KDJ93} to our geometry by creating a \textit{dark funnel} for the atoms. This serves to increase the atomic density in the dark region as the atoms approach the HCPCF. To set up the dark funnel, we replaced the Gaussian-shaped repumper beam by two orthogonally aligned beams with funnel-shaped opaque masks in their beam paths (see Fig.~\ref{fig:Fig1}(b)) and imaged them onto the cloud and fiber. Radiation resonant with the transition $|F=2\rangle\rightarrow |F'=2\rangle$ (see Fig.~\ref{fig:Fig1}(c)) propagated through the fiber to serve as a depumper, which confines the population mainly in state $|F=1\rangle$ \cite{APE94}. We chose the width of the funnel to be twice as wide as the 1/e full width of the depumper to avoid heating by repumper and depumper pumping cycles. Ideally, this should enable all atoms to reach the fiber tip simply by gravitational force. However, we observed that though the center of the cloud in the dark funnel moved down towards the fiber when we switched on the depumper, no atoms made it into the fiber. We attribute this to small amounts of repumper stray light which leads to non-vanishing intensity right above the fiber where the extensions of the dark funnel are very small. To overcome this problem, we employed a two-stage loading procedure (see timing in Fig.~\ref{fig:Fig2}(a)) : In a first step, we collected atoms in the (dark) MOT for 990~ms. We then shifted the center of the MOT down towards the fiber by a magnetic offset field and ramped up the quadrupole field gradient during the course of 30~ms. Finally, we held the atom cloud right above the fiber tip for an arbitrary duration. At this empirically found optimum position of the cloud there was significant overlap with the FORT and atoms were guided into the HCPCF.

To probe the atoms which made it into the fiber core, we sent a probe beam of circular polarization and resonant with the transition $|F=1\rangle \rightarrow |F'=2\rangle$ through the fiber. We spatially filtered the beam after the HCPCF with a single-mode fiber and detected the transmitted light with an avalanche photodiode (SAR500H1, Laser Components). This served to detect only light which propagated through the HCPCF core. To avoid probing also atoms which might still be located above the fiber, we ``purged" this region with a strong (Gaussian-shaped) repumper beam before each absorption measurement.

Figure~\ref{fig:Fig2}(c) shows the \textit{absolute} transmitted probe beam power without and with atoms loaded into the fiber. The probe beam is switched on at time $t=0~\mu$s. With no atoms present in the HCPCF we observed constantly high transmission (black squares). When atoms are loaded into the HCPCF (red circles), the probe beam is completely absorbed initially until it has optically pumped all atoms into state $|F=2\rangle$. During the pumping process, each atom absorbs on average two photons before ending up in state $|F=2\rangle$. The area below the difference signal (blue triangles) of transmitted probe power with and without atoms loaded into the fiber hence represents the energy absorbed by the atoms in the fiber. This permits determination of the atom number in the HCPCF. From the data shown in Fig.~\ref{fig:Fig2} we deduce a number of $N=2.5(3)\times10^5$ atoms loaded into the HCPCF. The corresponding loading efficiency from the MOT (with $N_{MOT}=1.0(2)\times10^7$ trapped atoms, as measured in the same fashion) into the HCPCF is $\eta=2.5(6)\%$. From a Monte-Carlo simulation of loading the FORT, taking a transverse extension of $5.5~\mu$m ($1/e$ full width) of the FORT and a trap depth of 5~mK into account, we deduced a temperature of $450~\mu$K and a transverse diameter of $1.4~\mu$m ($1/e$ full width) of the atom cloud inside the HCPCF. From these numbers we estimated the OD on the transitions $|F=1\rangle\rightarrow |F'=1,2\rangle$ for an unpolarized medium, yielding $OD\simeq1400$ \cite{BHP11}.

To determine the OD also experimentally, we recorded absorption spectra on the transitions $|F=1\rangle \rightarrow |F'=0,1,2\rangle$ with a weak probe beam ($P\simeq20$~pW) detected by a photon counter (PerkinElmer, SPCM AQRH-12). We then compared the measured spectra to calculations with the OD as fit parameter \cite{PWB12}. We prepared the atoms inside the HCPCF by the sequence described above, holding the atom cloud for 20~ms above the fiber for loading. Then, we modulated the FORT with a frequency of 250~kHz \cite{BHP11}, to provide time slots of 800~ns duration each, where the FORT was switched off for the absorption measurements (see Fig.~\ref{fig:Fig2}(b)). During each of these periods we measured the transmission of the probe beam during a gate time of 680~ns. The FORT was then switched on again to recapture the atoms. This allowed us to probe the same atoms up to 50 times without any significant losses.

Fig.~\ref{fig:Fig2}(d) shows a typical spectrum measured for $N=1.4(4)\times 10^5$ atoms inside the HCPCF. The experimental data (symbols) is accompanied by simulation results (lines) for different ODs. The best agreement between experiment and calculation (black line) is reached for $OD=(300(45),1000(150),1000(150))$ on the transitions $|F=1\rangle\rightarrow |F'=0,1,2\rangle$. To the best of our knowledge this represents the largest effective OD measured so far on a non-cycling transition, both for ultracold and warm gases. We note that the measured OD agrees well with the OD estimated from calculations \cite{BHP11} involving the measured number of atoms (which was varying within the several hours of measurement by $\pm4\times10^4$), estimated temperature of the atoms in the HCPCF according to the trap depth, and assuming equally populated Zeeman levels.

We note several issues: Comparison of experimental data and the results of a Monte-Carlo simulation of the loading procedure indicates that we have not yet reached the maximum loading efficiency. Losses of atoms from the FORT as they propagate through the HCPCF seem to set the limit at the moment. Moreover we found, that the loading efficiency also depends upon the longitudinal mode of the FORT. Single mode operation tends to produce lower efficiencies, whereas multi-mode operation increases the efficiency by a factor of about two. We are currently working on detailed studies and a characterization of the loading procedure to understand these features and possibly further improve the OD. The results will be subject of a future publication.

In conclusion, we demonstrated the up-to-date highest loading efficiency of laser-cooled $^{87}$Rb atoms from a MOT into a HCPCF of 7~$\mu$m core diameter. We loaded up to $N=2.5(3)\times10^5$ $^{87}$Rb atoms into the fiber with an efficiency of $2.5(6)\%$. By comparison of a measured absorption spectrum with calculated data, we determined the OD of our one-dimensional medium as 1000(150) on the transitions $|F=1\rangle\rightarrow |F'=1,2\rangle$. This represents the highest effective OD ever measured for ultracold atomic media on transitions relevant to light storage protocols and paves the way towards a new class of experiments requiring strong light-matter interactions in the field of quantum and nonlinear optics.

The authors thank M. Hain for assistance with the absorption measurements, L.P. Yatsenko, B.W. Shore, and R. Nolte for valuable discussions and the electronic materials group at Technische Universit\"at Darmstadt for the scanning electron micrographs of our HCPCF. The research leading to these results has received funding from the Deutsche Forschungsgemeinschaft and the European Union Seventh Framework Programme (FP7/2007-2013) under grant agreement n$^\circ$ PCIG09-GA-2011-289305.

%\bibliography{../EIT,../AtomTrapping,../BEC,../PBGF,../Entanglement,../TGGas}

%

\end{document}